\documentclass[12pt]{article}
\begin{document}
\begin{center}
{\em On Realism and Quantum Mechanics}\\
\vskip3mm
{\small Giuseppe Giuliani}\\
{ \small Dipartimento di Fisica `Volta', Via Bassi 6, 27100 Pavia, Italy}\\
{\small giuliani@fisicavolta.unipv.it}\\
{\small PACS 03.65.Bz}
\end{center}
\vskip5mm\par\noindent
{\bf Abstract.}
{\small A discussion of the quantum mechanical use of superposition or
entangled states shows that descriptions containing only statements
about state vectors and experiments outputs are the most suitable
for Quantum Mechanics. In particular, it is shown that statements
about  the undefined values of physical quantities before
measurement can be dropped without changing the predictions of the
theory. If we apply these ideas to {\em EPR} issues, we find that
the concept of non~-~locality with its `instantaneous action at a
distance' evaporates. Finally, it is argued that usual treatments of
philosophical realist positions end up in the construction of
   theories whose major role is that of
being disproved by experiment. This confutation  proves simply that
the theories are wrong; no conclusion about realism (or any other
philosophical position) can be drawn, since experiments deal always
 with theories and these   are never logical consequences of
 philosophical positions.}
\vskip5mm
\section{Introduction}
  The question of the (im)possible coexistence between realism and Quantum Mechanics
 goes back to the birth of the latter;
 however, the addressed issues and the relevance given to them have
 changed since then. The debate has shifted from the  discussion
 of the epistemological status
 of Quantum Mechanics to specialized
 topics connected with the so called {\em EPR} paradox \cite{ref:epr}: the turning
 points have been the paper by Bell on his inequalities \cite{ref:bell} and the
 Aspect's experiment on {\em EPR} correlated  photons pairs \cite{ref:aspect}.
\par
 Fuchs and Peres hold that
`Quantum Mechanics needs no interpretation'  \cite{ref:nointer}.
This statement  is equivalent to say, as we do in the abstract, that
`descriptions containing only statements about state vectors and
experiments outputs are the most suitable for Quantum Mechanics'. In
the following, we shall use the term `interpretation' in two senses.
Given a physical description based on a set of equations, there is a
`minimal (and necessary) interpretation': it tells us which are the
procedures for measuring (at least some of) the physical quantities
that appear in the equations.  The usual interpretation of Quantum
Mechanics is based on a `second level' interpretation characterized
by the assumption  that `in some instances, the physical quantities
of a system do not have a definite value before measurement'. It
will be shown that this assumption can be dropped without changing
the predictive content of Quantum Mechanics.
\par
The philosophical background of the main thesis sustained in this
paper may be defined as a `tempered' realist one. However, since the
thesis of this paper, though suggested by this philosophical
position, {\em does not logically depend on it}, we omit its
illustration. The interested reader can see  \cite{ref:quale}.
   \section{Realism and Quantum Mechanics\label{rqm}}
  Since Quantum Mechanics makes statements about the World, it is a
    realist description in the
     following sense:
\begin{itemize}
    \item [$\natural$] it is interpreted on the basis
    of, at least,
    the basic realist assumption: `there is an external
     World whom the observer belongs to';
    \item [$\natural$] it describes
      experiments and uses experimental reports: descriptions and reports
      of experiments are realist discourses.
\end{itemize}
\subsection{The superposition state\label{super}}
 As a basic case,  let us consider a
two states system $S$ described by the state vector:
\begin{equation}\label{duestati}
    |\psi\hspace{-4pt}>\:= \frac{1}{\sqrt{2}}\,(|\psi_1\hspace{-4pt}>+|\psi_2\hspace{-4pt}>)
\end{equation}
where $1$ and $2$ label the two states of $S$ (the factor
$1/\sqrt{2}$ implies that the two states are equally probable). Let
us further suppose that the eigenvalue of a physical quantity $A$ of
the system described by the eigenvector $|\psi_1\hspace{-4pt}>$ is
$a_1$ and, correspondingly, $a_2$ for $|\psi_2\hspace{-4pt}>$.
\par
 Here are two
possible interpretations of equation (\ref{duestati}):
\begin{enumerate}
    \item [($S1)$]  The system $S$ is {\em described} by the state vector (\ref{duestati}).
    If a measure of the physical quantity $A$ is made on the
     system $S$,
    then the probability of finding $a_1$ or $a_2$ is $1/2$.
    This is a `minimal' interpretation.
    \item[($S2$)]\label {S2}  The system $S$  is  {\em described} by
the
  state vector (\ref{duestati}).
    The physical quantity $A$ does possess, before the
measurement, neither the value $a_1$ nor the value $a_2$; during the
measurement, the system passes from the `superposition state' to the
state $|\psi_1\hspace{-4pt}>$ or $|\psi_2\hspace{-4pt}>$ with the
associated properties [statement $(M2)$: from `The physical quantity
$A$\dots' to the the end]. $(M2)$ is a realist assertion because it
affirms something about the World; moreover, since it is implicitly
considered as experimentally not testable, it is a  {\em
metaphysical} statement. This interpretation is of `second level',
because of  statement $(M2)$.
\end{enumerate}
The predictions of Quantum Mechanics depends {\em only} on the form
of the state vector (\ref{duestati}). ($M2$)
 is not used in
the deduction chain that leads to experimentally testable
predictions: {\em it can be dropped without changing these
predictions}. As a matter of fact $(M2)$ is used only in the
interpretation of experimental results. $(S2)$ leads directly to
  Schr\"odinger's `cat
paradox', to all its variations and extravagant implications. With
$(S1)$, instead, we shall avoid any trouble.
\par
     As a second  example, let us  consider a beam of linearly
polarized light coming out from a polaroid whose axis is, say, along
the $x$ axis. If the beam is falling on a second polaroid whose axis
$x'$ is tilted by an angle $\theta$ from that of the first polaroid,
Quantum Mechanics predicts that each photon has  probability
$\cos^2\theta$ of passing through the second polaroid \footnote{Also
classical electromagnetism yields the same prediction if we assume
that the probability of traversing the second polaroid is
proportional to the classical predicted intensity of the light
passing the second polaroid. Similarly, in the case of the two slits
experiment, classical electromagnetism can predict what is the
probability for a photon to reach a point on the screen, if it is
assumed that this probability is proportional to the classical light
intensity predicted for that point. Of course, this is an {\em ad
hoc} adjustment of Maxwell's theory; however, it is conceptually
interesting. For a detailed discussion, showing also that even
classically it is not possible to say which is the energy path
between the slits and the screen, one might see \cite{ref:ggib}.}.
Quantum Mechanics does so by describing the
 incoming photon by a `superposition'
of linear polarizations along two perpendicular directions $x',y'$
\footnote{Also classical electromagnetism does, of course, a similar
thing with the electric field.}:
\begin{equation}\label{secondopola}
    |x>= |\,x'>  \cos\theta + |\,y'> \sin\theta
\end{equation}
 The experiment can be described in the following way.
The photon impinging on the second polaroid {is}
    linearly polarized along $x$ since it has passed the first
    polaroid. This statement derives from the   operational
    definition according to which
     a photon is said to be `linearly polarized along $x$'
     if it  has passed a polaroid oriented along $x$.
     Taking into account the type of experimental
    apparatus, we {\em describe} the photon by equation (\ref{secondopola});
    the photon coming out from the first polaroid has
probability $\cos^2\theta$ of passing through the second one. {\em
The interaction between the photon and the polaroid changes the
polarization of the photon}.
\par
On the basis of equation (\ref{secondopola}), no one would hold that
the photon does not possess a definite polarization before the
second measurement (we know  it is polarized along $x$):
nevertheless, we describe the photon by a `superposition' state
vector. This means that the photon under study can be described by a
superposition state vector even when it has a definite value of the
polarization.
     Equation (\ref{secondopola}) is
clearly written down by having in mind, as in the classical
treatment, the type of experiment we are performing (the impact with
the second polaroid).
           \subsection{{EPR} issues} The
debate addresses (at least) four issues: realism, locality,
causality and completeness of Quantum Mechanics. Historically, the
starting point has been the problem of completeness of Quantum
Mechanics, posed by the already quoted paper by Einstein, Podolsky
and Rosen \cite{ref:epr}. In recent years the attention has been
focused on $EPR$ type experiments.
\par
 If we
picks up a typical contemporary paper concerning {\em EPR}
arguments, we are
 facing the following situation: the orthodox
description is presented as a `non~-~realist' one; philosophical
realist positions are {\em translated} into a physical theory whose
predictions are necessarily different, at least in some cases, from
the ones of Quantum Mechanics; fatally, the `realist' theory is
disproved by experiments. In these papers, realist positions are
characterized, among other conditions, by the statement ($SR$) that
`it is possible to attribute a definite value to a physical quantity
of a system before  measurement'. This characterization is untenable
because it is based on the assumption that $(SR)$ is a philosophical
assertion; as we shall see, it is a physical assertion and,
therefore, it can be tested~-~at least in principle~-~by experiment
(section \ref{twopossible}).
\subsubsection{\em Photon pairs correlated in polarization\label{correlati}}
 Nowadays, these experiments are carried out by using
photons pairs produced by parametric down conversion \footnote{See,
for instance,  \cite{ref:pdc} and references therein.}. However, we
shall discuss first the `classical method' based on pairs produced
by atoms in a cascade process \footnote{The first {\em EPR} type
measurement with photons pairs produced by a cascade emission is due
to Kocher and Commins \cite{ref:kocher}. The photons of the pairs
produced by this type of source are not correlated in direction:
therefore, the number of detectable correlated pairs is greatly
reduced. }.
\par
 Let us consider a pair of photons produced by a
single atom  and flying away in opposite directions (for instance
$\pm z$).
 When the
photons are well apart, the photon $\nu_1$ flying, say,  along $z$
is analyzed by  polaroid $A$ while the other photon $\nu_2$ (flying
along $-z$) is analyzed by  polaroid $B$; behind the polaroids there
is a photon detector \footnote{Two filters, one on the $z$ path and
the other on the $-z$ path, block the `wrong'
photons.}$^,$\footnote{Starting from Aspect's experiment
 \cite{ref:aspect}, the polaroids have been replaced by birefringent
analyzers; however, this complication, suggested by Bell's type
inequalities known as $BCHSH$ \cite{ref:bchsh}, can be avoided here
since we are not interested   in hidden variables theories. }. $A$
and $B$ make many measurements: we would like to know which is the
correlation between the measurements of $A$ and $B$ as a function of
the angle $\theta$ between their  axis.
\par
 The photon pair is described by the state vector:
\begin{equation}\label{circolare}
|\psi(\nu_1,\nu_2)\hspace{-4pt}>\:=\frac{1}{\sqrt{2}}\,(|R_1,R_2\hspace{-4pt}>
+\,|L_1,L_2\hspace{-4pt}>)
\end{equation}
where  $(|R\hspace{-4pt}>, \,|L\hspace{-4pt}>)$ are circular
polarizations states. However, since a circularly polarized photon
can be described as a combination of two linear polarizations, the
above equation can be written in the form:
\begin{equation}\label{corre}
|\psi(\nu_1,\nu_2)\hspace{-4pt}>\:=
\frac{1}{\sqrt{2}}\,(|x_1,x_2\hspace{-4pt}>+\,|y_1,y_2\hspace{-4pt}>)
\end{equation}
where  $(|x>, \,|y>)$ are linear polarization state vectors.
\par
 Let us now suppose that the measurement by $A$
is made before the one made by $B$ \footnote{We are following
 the treatment by Aspect  \cite{ref:aspect_2}.}.
 If the photon pair is described
by (\ref{circolare}), then the probability that photon $\nu_1$
passes through  $A$ is $(1/2)$;  if $\bf a$ is the direction of the
axis of $A$, the photon pair, after the measurement made by $A$, is
described by the state vector:
\begin{equation}\label{dopoa}
    |\psi'(\nu_1,\nu_2\hspace{-4pt}>= |\bf a,a\hspace{-4pt}>
\end{equation}
Therefore, if polaroid $B$ is oriented as polaroid $A$, photon
$\nu_2$ passes through $B$; if, instead,  $B$ is tilted by the angle
$\theta$ with respect to  $A$, photon $\nu_2$ will pass through $B$
with probability $\cos^2\theta$ (Malus law). Then
 the probability  that
photon $\nu_1$ passes through $A$ and photon $\nu_2$ passes through
$B$ is given by:
\begin{equation}\label{qmab}
    P(A,B)=\frac{1}{2}\cos^2\theta
\end{equation}
This equation can, of course, be derived directly from (\ref{corre})
by calculating the probability that photon $\nu_1$ passes through
$A$ and photon $\nu_2$ passes through $B$ without considering the
details of the experiment.
\par The usual interpretation is:
\begin{enumerate}
\item [$(QM_{1})\, A$] Before the measurement,
     the photons of each pair do not posses
    a definite value of the polarization.
    \item [$(QM_{1})\, B$] {\em Therefore}, the  photons pairs produced
    by
the source
    are { described} by the state
    vector (\ref{circolare}).
              \item[$(QM_{1})\, C$] The probability for $\nu_1$
     of passing through
    $A$ is $1/2$. If photon $\nu_1$ passes through
    $A$, then photon $\nu_1$ is (operationally) linearly polarized
    along the direction of the axis of $A$.  Contemporaneously,
    photon $\nu_2$ {\em assumes} the same polarization.
   \item [$(QM_{1})\, D$]
    If photon $\nu_1$ passes through $A$, then photon $\nu_2$ will pass through $B$ with probability $\cos^2\theta$,
    where $\theta$ is the angle between the axis of the two
    polaroids (Malus law). Therefore, the probability that photon
    $\nu_1$ passes through $A$ and photon $\nu_2$ passes through $B$
    is $(1/2)\cos^2\theta$.
  \end{enumerate}
Let us now consider the following description obtained from the
previous one by omitting the assumption $(QM_{1})\, A$:
            \begin{enumerate}
    \item [$(QM_{0})\, A'$]The pair of photons produced by the source
    {\em is described} by the state vector (\ref{circolare}).
        \item[$(QM_{0})\, B'$]  The probability for $\nu_1$
     of passing through
    $A$ is $1/2$. If  photon $\nu_1$ passes through
    $A$, then the photon pair {\em is described} by  state vector
    (\ref{dopoa});
        therefore, photon $\nu_2$ will pass through  $B$
    if it is oriented as  $A$.
   \item [$(QM_{0})\, C'$]
    If photon $\nu_1$ passes through $A$,
     then photon $\nu_2$ will pass through $B$ with probability $\cos^2\theta$,
    where $\theta$ is the angle between the axis of the two
    polaroids (Malus law). Therefore, the probability that photon
    $\nu_1$ passes through $A$ and photon $\nu_2$ passes through $B$
    is $(1/2)\cos^2\theta$.
  \end{enumerate}
    The  difference between the two descriptions is due to the
  presence~-~in the usual one~-~of the statement $(QM_{1})\, A$:
  it entails,  in sentence
$(QM_{1})\, C$, that
  a definite value of the polarization is given
to the photon $\nu_1$ by  measurement; as a consequence, it is held
that `contemporaneously,
    photon $\nu_2$ {\em assumes} the same
     polarization'. This sentence
   incorporates an `instantaneous action at a distance'. It is claimed that
         special relativity
        is not violated since there is no information transport
        between
        $A$ and $B$: we can verify the correlations between the
        measurements made by $A$ and those made by $B$ only by
        bringing together their data. However, the process under
        challenge is not the reading by a human observer (who
        collects the data of $A$ and $B$) but the purported
        physical process according to which the polarization
        measured  by $A$ on photon $\nu_1$ is instantaneously
        `transmitted' to photon $\nu_2$: within the usual        description,
        without this `transmission', the
        human observer who collects the data from $A$ and $B$ would never
        see the observed correlations. As a matter of fact,
        given the unpopularity of the instantaneous action at a distance, it is spoken
        of {\em non~-~local} effects. Non~-~locality with its associated
instantaneous action at a distance is a direct consequence of
statement $(QM_{1})\, A$ about the undefined value of polarization
before measurement. If we drop this assumption, we switch to the
labeled description that  deals only with state vectors and
experimental
  outputs and avoids non~-~local effects. However, in the labeled
  description, {\em the choice of the starting state vector needs to be
  justified}. This will be done in next section.
\subsubsection{\label{opcp}\em An operational definition of `correlated polarization'}
The primed description of the Aspect experiment
 suggests the following operational definition of {\em correlated
polarization} of the photon pairs:
\begin{quote}
Put the two polaroids with their axis parallel. {\em If}, for every
photon pair,  photon $\nu_1$ passes through  $A$  {\em and} photon
$\nu_2$ passes through $B$, {\em then} the twin photons are said to
be correlated in polarization.
 \end{quote}
This definition, based on  state vector  (\ref{circolare}) (or,
equivalently (\ref{corre})),  correlates an empirical property of
the photon pairs with the state vector that describes them. In
section \ref{bellstates}, where all the so called `four Bell states'
are considered, it will be shown that to every `Bell state' (state
vector) corresponds a uniquely defined `correlated polarization' of
the twin photons. Therefore, a one~-~to~one correspondence exists
between state vectors and `correlated polarization' of the photon
pairs. Since photon pairs described by different Bell states have
different `polarization correlations', we must conclude that this
`polarization correlation' is a property of the photon pairs as they
are produced by the source. In next section, the `polarization
correlation' of photon pairs produced by a cascade process will be
further analyzed on the basis of two possible experiments.

\subsubsection{\em Two possible experiments on
the polarization of the twin photons beams of the Aspect
experiment\label{twopossible}} The `polarization correlation'
operationally defined in previous section (and extended to all `Bell
states' in section \ref{bellstates}) deserves a deeper
understanding.
\par
 Our acquired
knowledge says that conservation laws are valid in every single
atomic event. Therefore, since the twin photons are emitted by two
consecutive transitions from an initial $J=0$ ($^1S_0$) to an
intermediate $J=1$ state ($^1P_1$) and to a final $J=0$ state
($^1S_0$),  the twin photons {\em are expected} to be both right or
left circularly polarized  (angular momentum conservation).
Therefore:
\begin{enumerate}
  \item [a)] The beams flying along  $ \pm z$ should be made up by a
(statistically) equal number of right and left circularly polarized
photons:  the beams should be unpolarized.
  \item [b)]  However, each photon {should} have a
definite polarization: right or left.
\end{enumerate}
  This point is usually
neglected, probably owing to the idea that, before measurement, the
twin photons cannot have a definite polarization since they are
described by a pure state. This objection notwithstanding,  let us
follow this line of reasoning and see if it can deepen our
understanding of the `correlated polarization' property.
\par
As suggested by classical optics, points a) and b) can be
experimentally checked by the combined use a quarter wavelength
plate and a polaroid. It is to be stressed that these measurements
tell us which is the polarization of the photons  before their
entrance  in the experimental apparatus (plate+polaroid).
\begin{quote}
a)   Each photon of the beam should be either right or left
polarized
 with probability
$1/2$. If it is right, it will be, after the plate, linearly
polarized along a direction $\bf a$ tilted by an angle $\pi/4$ with
respect to the optical axis of the plate \footnote{In literature,
there are two different definition of right/left polarization: which
is right for one, is left for the other; and viceversa. Of course,
our argumentation is valid whichever definition is chosen. }. Then,
it will have a probability $\cos^2\theta$ of passing through a
polaroid whose axis makes an angle $\theta$ with $\bf a$. Then the
probability of passing the polaroid is $(1/2)\cos^2\theta$. \par
 If the photon
is left polarized, it will be, after the plate, linearly polarized
along a direction $\bf b$ tilted by an angle $-\pi/4$ with respect
to the optical axis of the plate.  Then, it will have a probability
$\sin^2\theta$ of passing through a polaroid whose axis makes an
angle $\theta$ with $\bf a$. Then the probability of passing the
polaroid is $(1/2)\sin^2\theta$. \par Since each photon is either
right or left polarized, the probability of passing through the
polaroid is $(1/2)\cos^2\theta+(1/2)\sin^2\theta=1/2$, independent
of $\theta$. Therefore,  the intensity of the light beam does not
change by rotating the polaroid.
\par
b)
 A quarter wavelength plate and a polaroid (with its
axis tilted by $\pi/4$ with respect to the optical axis of the plate
in order to detect, say, right circularly polarized photons) are
inserted into the photons' $\pm z$ paths  (with the blocking
filters). We should observe that when a photon is detected along $z$
a photon is detected also along $-z$: then, we must conclude that
the twin photons were right circularly polarized {\em before} their
entrance into the measuring apparatus (plate plus polaroid). By
rotating both polaroids by $\pi/2$, the photomultipliers will detect
in coincidence only left circularly polarized photons. Finally, by
putting the two polaroids in such a configuration that the two
measuring devices (plate + polaroid + photon detector) detect,
respectively, right (along $z$) or left (along $-z$) circularly
polarized photons, we shall have a `click' only along $z$ or $-z$
for every photons pair.
\end{quote}
\subsubsection{\em A third, possible description\label{third}}
Let us suppose that the basic idea leading to the experiment
proposed in b) is correct: the twin photons are both right or left
polarized. Then, using the definition of `correlated polarization'
of section \ref{opcp}, we can build up the following description of
the experiment:
\begin{enumerate}
    \item [$(QM_{0})\, A''$]The polarization  of the photons of a pair
     produced by the source is correlated.
    \item[$(QM_{0})\, B''$]Each photon of
    the pair
  does  possess a definite value of the polarization: it is
    either right or left circular.
    \item[$(QM_{0})\, C''$]Since photon $\nu_1$ is right or left circularly polarized,
    it has probability $1/2$ of passing through polaroid $A$.
        \item[$(QM_{0})\, D''$]If photon $\nu_1$ passes through
    $A$, {\em since the polarization of the photon
    pair is correlated}, then photon $\nu_2$  passes through
    $B$ if its orientation is the same as that of $A$; otherwise,
    it will pass through $B$ with probability $\cos^2\theta$,
    where $\theta$ is the angle between the axis of the two
    polaroids (Malus law). Therefore, the probability that photon
    $\nu_1$ passes through $A$ and photon $\nu_2$ passes through $B$
    is $(1/2)\cos^2\theta$.
    \end{enumerate}
This description is based on two features that are usually
considered as incompatible: pure states and definite values of a
physical quantity before measurement. However, the fact that it
leads to the correct predictions suggests that this incompatibility
may not be there: only the experimental test suggested above could
clear this point.
\subsubsection{\em Extension to photons pairs produced by parametric
down conversion\label{bellstates}}
 Parametric down conversion allows to produce all of the so called
 `four Bell states':
 $$\psi^{\pm}=
 (1/\sqrt{2})(|x_1,y_2\hspace{-4pt}>\hspace{-2pt}\pm\hspace{2pt}
 |y_1,x_2\hspace{-4pt}>);
 \qquad \phi^{\pm}=
 (1/\sqrt{2})(|x_1,x_2\hspace{-4pt}>\hspace{-2pt}\pm \hspace{2pt}|y_1,y_2\hspace{-4pt}>)$$
  where $x$ and $y$ are two perpendicular
 linear polarization states.
A  general
  definition of `correlated polarization', applicable
  to each of the four Bell states, is as follows:
  \begin{quote}
   Put the two measuring polaroids $A$ and $B$ in such a configuration (relative
 orientations of their axis) that, for every photon pair, if a photon of the pair passes
 through $A$, then the twin photon passes through $B$. If this is possible,
  then the photons pairs
 are said to be correlated in  polarization.
  \end{quote}
  The relative orientation of the two polaroids depends on which of
 the four `Bell states' is considered:
 parallel for the $\phi^+$ state; perpendicular for the $\psi^+$ state;
 $\theta_B=-\theta_A$ for the $\phi^-$ state; $\theta_B=-\theta_A+\pi/2$ for the
 $\psi^-$ state.
 Using the above general definition of `correlated polarization', it is possible to set up a
 description analogous to the doubled primed one discussed in section \ref{opcp}
 for
 each of the four Bell states.
 \begin{quote}
    Example. In the case of the $\phi^-$ state, the double primed
    description goes as follows: i) the polarization of the photons
    of a pair is correlated; ii) before any measurement, each photon of
    the pair
  does  possess a definite value of the polarization: it is
    either $x$ or $y$ linear polarization;
 iii)   since photon $\nu_1$ is either $x$ or $y$ linearly polarized,
    it has probability $1/2$ of passing through polaroid $A$
    oriented along $\theta_A$ (arbitrarily chosen);
     iv)   if photon $\nu_1$ passes through
    $A$, since the polarization of the photons
    of the pair is correlated, then photon $\nu_2$  passes through
    $B$ if it is  oriented along $\theta_B=-\theta_A$; otherwise,
    it will pass through $B$ with probability
    $\cos^2(\theta_A+\theta_B)$ (Malus law); v) therefore, the probability that
    photon $\nu_1$ passes through $A$ and photon $\nu_2$ passes
    through $B$ is given by $(1/2)\cos^2(\theta_A+\theta_B)$.
    \end{quote}
         \section{Causality\label{causa}}
 The causality principle has been frequently challenged after the
 formulation of Quantum Mechanics.
 It is
 claimed that, while probabilistic theories of classical physics
 reflects our ignorance about phenomena, the probabilistic nature
 of Quantum Mechanics  reflects the indeterministic nature of
 quantum phenomena.
 \par
 The attribution  of a general
 feature of a theory (in this case its probabilistic nature) to the World
 constitutes a strong {\em realist}  assertion about how things
 behave
  in the
 World: then, we are called to carefully evaluate its plausibility.
 However, the main point is that the `causality principle',
 understood as a methodological commitment to the searching for causes, has
 been one of the propulsive forces  of scientific knowledge:
  a discipline that, on the
 basis of hardly conclusive evidence, is {\em really} abandoning this  commitment,  is
 doomed to drain its vital sources.

             \section{Conclusions}
The debate about $EPR$ issues has been characterized by the
confrontation of a `realistic stand'~-~originally attributed to
Einstein~-~with a `quantum mechanical' one. This confrontation has
been played, though not uniquely,  about  the question {\em `Q'}:
``Can we attribute a definite value to a physical quantity before
measurement? Or is  the process of measurement that attributes this
definite value to the quantity under measurement?'' The answer given
in this paper is:
\begin{enumerate}
  \item [A)] Question {\em `Q'} is a physical question, not a philosophical one.
   As such, it can be
  decided by experiment (section \ref{twopossible}).
  \item [B)]In the case of $EPR$ experiments with twin photons pairs the
   state vectors describing the photon pairs are chosen
  because they describe the correct correlation between the
  polarizations of the twin photons and not on the basis of the
  defined/undefined value of the polarization before measurement
  (sections \ref{opcp} and \ref{bellstates}).
  \item [C)] The non~-~locality effects appearing in the usual
  interpretation of these $EPR$ experiments are due to the
  hypothesis (explicit or implicit) that `only the process of measurement
 attributes a definite value to the quantity under measurement'
  (section \ref{correlati}). If
 we drop this hypothesis, and this can be done without changing the
 predictions of the theory, non~-~locality effects disappear
 (sections
 \ref{correlati} and \ref{third}).
 \end{enumerate}
The discussion about the epistemological status of Quantum Mechanics
has been characterized, since its beginnings, by the confrontation
of different philosophical positions. This discussion has
increasingly spread over and outside  the scientific community and
has produced an impressive huge number of papers and books: it is
hard avoiding the uncomfortable feeling that this discussion has
drained much more intellectual resources than deserved. \par
 Starting from
Bell's paper  \cite{ref:bell}, a bizarre game has been
 played: to build up theories labeled as realist; and
to realize increasingly sophisticated experiments in order to
disprove them \footnote{However, these sophisticated experiments
have enlarged our knowledge on how to produce and manipulate {\em
correlated} photons pairs.}. If we like, we can go on playing the
game; however, since the postulates of a theory cannot be logically
deduced from a philosophical position, we must be aware that {\em
experiments disprove sometimes a theory but never a philosophy}.
\vskip3mm\par\noindent
{\small {\bf Acknowledgements.}
  I had the opportunity of discussing the content
of this paper  during talks and  informal meetings and by email.
Thanks to all those who, with their comments and/or objections, have
contributed to clarify some basic points. Particular
 thanks are due to Giancarlo Campagnoli and Peppino Sapia
for friendly discussions and critical reading of the first version of
the manuscript.}

\end{document}